\documentstyle[preprint,prl,aps,psfig]{revtex}
\begin{document}
\draft
\title{Skyrmions in the Fractional Quantum Hall Effect}
\author{R.K. Kamilla, X.G. Wu, and J.K. Jain}
\address{Department of Physics, State University of New York
at Stony Brook, Stony Brook, New York 11794-3800}
\date{Feb. 16, 1996}
\maketitle
\begin{abstract}

It is verified that, at 
small Zeeman energies, the charged excitations in the 
vicinity of 1/3 filled Landau level are skyrmions 
of composite fermions, analogous to the skyrmions of electrons 
near filling factor unity. These are found to be relevant, 
however, only at very low magnetic fields.  
\footnotetext{Accepted for publication in {\it Solid State Communications}
 (1996).}
\end{abstract}

\subsection{Introduction}

Since the discovery of the fractional quantum Hall effect (FQHE) \cite
{Tsui}, the nature of various states of interacting electrons confined to 
their lowest LL has been a subject of great interest. Recently, a new
theory is being explored according to which the system of strongly
correlated electrons at magnetic field $B$ maps on to a system of
weakly interacting composite fermions at an effective magnetic field
$B^*=B-2m\rho\phi_0$, where $m$ is an integer, $\rho$ is the electron
density, and $\phi_0=hc/e$ is the quantum of flux \cite {Jain}. 
The wave functions for the composite fermions are constructed by
multiplying the wave functions of weakly interacting electrons at $B^*$
by the Jastrow factor $\prod_{j<k}(z_j-z_k)^{2m}$, where
$z_j=x_j+iy_j$ denotes the position of the $j$th electron \cite {Jain}. 
This simple theory successfully provides
a detailed and unified description of various liquid states of
electrons in the lowest Landau level (LL), in particular, of 
spin-polarized incompressible FQHE states and their charged and
neutral  excitations \cite {Jain,Dev}, the compressible 
state at and near 1/2 filled LL \cite {HLR}, 
and incompressible odd-denominator FQHE states at small Zeeman 
energies \cite {Du,Wu}. The last case is of relevance since the Zeeman
energy is quite small in GaAs, and there exists good experimental
evidence that many FQHE states in low-density samples are not fully
polarized \cite {Du,Eisenstein}. The ground state at 1/3 is, however, fully
polarized even in the absence of Zeeman splitting, and is well described 
by the Laughlin wave function \cite {Laughlin}, which is interpreted
in the composite fermion (CF) scheme as one filled LL of composite fermions.
Nakajima and Aoki \cite {Nakajima} have shown that the (neutral) 
spin-wave excitation at filling factor $\nu=1/3$ can be
quantitatively understood as the
spin-wave excitation of composite fermions at CF filling factor 
$\nu^*=1$. This Communication demonstrates analogous behavior  
for the (charged) skyrmion excitation near 
the 1/3 filled LL at zero Zeeman energy. A `hard-core' trial
wave function is found to be rather accurate, and allows 
an estimation of the number of reversed spins in the CF-skyrmion 
as a function of the Zeeman energy.

It is convenient to use a spherical geometry \cite {Haldane}, 
in which $N$ electrons move on the
surface of a sphere under the influence of a radial magnetic field.
For $N_\phi$ flux quanta threading the surface of the sphere,
the degeneracy of the
lowest LL is $2(N_{\phi}+1)$, where the factor of 2 originates
from the electron spin. According to the composite
fermion theory, interacting electrons at flux $N_{\phi}$  resemble 
weakly interacting composite fermions at flux 
\begin{equation}
N_{\phi}^*=N_{\phi}-2(N-1)\;\;.
\label{cfmap}
\end{equation}
The 1/3 state occurs at $N_{\phi}=3(N-1)$, which, relates to 
composite fermions at $N_{\phi}^*=N-1$, corresponding to a CF filling
factor $\nu^*=1$.  

The quasiparticle (quasihole) of the $\nu=1$ or 
state is obtained by decreasing (increasing) the flux $N_\phi$ by 
one unit. At large Zeeman energies, the quasiparticle (QP)  
involves reversal of a single  spin, and the quasihole (QH) state
is fully polarized. However, the spin configuration at small 
Zeeman energies is rather unusual, involving a reversal of possibly a
large number of spins
relative to the $\nu=1$ state \cite {Rezayi}. Such a quasiparticle
or quasihole is generically called a (QP- or QH-) skyrmion 
\cite {Kane,Sondhi}.

\subsection{$\nu=1$ skyrmion}

Let us first set the Zeeman energy equal to zero. 
Then, the total angular momentum $L$ and the total spin $S$ are good
quantum numbers. 

At $\nu=1$ ($N_{\phi}=N-1$), the ground state is fully polarized, as
one might have anticipated from Hund's first rule. 
However, the Hund's rule is maximally
violated just one flux quantum away, at $N_{\phi}=N$ (or $N_{\phi}=N-2$,
which is related to  $N_{\phi}=N$ by an exact
particle-hole symmetry in the lowest LL); here the ground state has
total spin $S=0$ and the total orbital angular momentum $L=0$ \cite
{Rezayi}.  In fact, the spectrum here contains a low energy branch of 
states with quantum numbers $(L,S)=(0,0)$, (1,1), ...  
$(\frac{N}{2},\frac{N}{2})$ (with $N$ taken to be an even 
integer for convenience), satisfying the property $L=S$ \cite
{JW,He,MacDonald}. The existence of this `skyrmion' branch can be 
understood as follows. Consider a hard-core delta
function interaction.  The states in which electrons
completely avoid one another irrespective of their spin will have zero
interaction energy, and will be referred to as `hard-core' states.
It has been known from numerical diagonalization \cite
{Rezayi,JW}, and also proven analytically \cite {MacDonald}, 
that for $N_{\phi}=N$, the hard-core states occur at precisely the
quantum numbers given above. 
Further, it has also been confirmed numerically that the low-energy 
eigenstates of the Coulomb interaction
are indeed almost identical to the hard-core states \cite
{JW} (although the longer range part of the Coulomb interaction
removes their degeneracy). 

The important question then is: What is 
the energy ordering of these states for the Coulomb interaction? 
The Hund's rule answers this question successfully, {\em provided it
is applied to composite fermions rather than electrons}.
Repeated applications of the CF
transformation, which progressively zooms into the lower energy
states, shows that the energy {\em increases} with $S$, consistent
with exact diagonalization results\cite {JW}. 

The hard-core wave function of the $\nu=1$ skyrmion 
is determined completely by symmetry (i.e., is independent of the
strength of the delta function interaction), since there is only one
hard-core state for any given set of $L$, $S$, $L_z$ and $S_z$ quantum
numbers.  We proceed to study large systems using the 
the hard-core wave function for the QH-skyrmion proposed by MacDonald, 
Fertig, and Brey \cite {MacDonald}. For $R$ reversed spins (relative to
the fully polarized quasihole), this wave function is given by 
\begin{equation}
\Psi_{SK}(R)=[\sum_{i_1,...,i_R}(z_{j_1} ... z_{j_{N-R}})
(\downarrow_{i_1} ... \downarrow_{i_R} \uparrow_{j_1}...
\uparrow_{j_{N-R}})] \Phi_1
\label{mfb}
\end{equation}
where $\uparrow$ and $\downarrow$ denote up and down spins, 
$j$'s denote particles
other than $i_1, ...,i_R$, and $\Phi_1$ is given by
\begin{equation}
\Phi_1=\prod_{j<k}(z_{j}-z_{k})\exp[-\frac{1}{4}\sum_i|z_i|^2]\;.
\end{equation}
$\Psi_{SK}$ has a well defined $S$ in the limit of large $N$ \cite
{MacDonald}.
For the spherical geometry, the product $z_{j_1} ... z_{j_{N-R}}$ 
is replaced by 
$v_{j_1} ... v_{j_{N-R}} u_{i_1}...u_{i_R}$, where $u$ and $v$ are
spinor coordinates \cite {Haldane}, and 
\begin{equation}
\Phi_1=\prod_{j<k}(u_iv_j-v_iu_j)\;,
\end{equation}
which is the wave function of the lowest filled LL in the spherical
geometry.

We have computed the Coulomb energy of $\Psi_{SK}(R)$  
for several values of $N$ by
Monte Carlo, shown in Fig.~1 \cite {Girvin}. 
The results are reasonably close to the thermodynamic limit, 
especially for the energy differences between different $R$ (for small
$R$).  The energy of the 50 electron system is
well approximated by (in units of $e^2/\epsilon l$)
\begin{equation}
E_{1}^{(0)}(R)=0.313+0.23\exp(-0.25 R^{0.85})\;.
\label{nu=1}
\end{equation}
It approaches $0.313 e^2/\epsilon l$ in the limit of large $R$, in
agreement with the result of Ref.~\cite {Sondhi}. 

At strictly zero Zeeman splitting,
the skyrmion ground state is a spin singlet, i.e., it involves a
spin-reversal for half of the electrons. However, this limit is
academic, since, even for a very small Zeeman energy, the lowest
energy state of the  
skyrmion has only a finite number of reversed spins, as anticipated
theoretically by Fertig {\em et al.} \cite {Fertig} and confirmed 
in several recent  experiments 
\cite {Barrett,Schmeller,Goldberg}. In the presence of a finite Zeeman
energy, a term $g^{*}\mu_B B R$ \cite {comm2}
must be added to $E^{(0)}(R)$ to get the full energy $E(R)$.
For parameters appropriate to GaAs
(Lande-$g$ factor $g^*=0.44$, effective electron mass $m^*=0.067$ 
$m_e$), and converting all energies into
Kelvin ($g^*\mu_B B \approx 0.30 B[T]\;K$, $e^2/\epsilon l\approx 50
\sqrt{B[T]}\;K$, $\hbar\omega_c \approx 20 B[T]\;K$, 
$B[T]$ measured in Tesla),
the total energy of the skyrmion  is given by
\begin{equation}
E_{1}(R)=[15.7 + 11.5 \exp(-0.25 R^{0.85})] \sqrt{B}+0.30 BR\;.
\end{equation}
The magnetic field above which the
minimum occurs at $R=0$ (i.e., the quasihole is fully polarized), 
estimated by the condition $E(0)=E(1)$, is
$B\approx 72T$. In general, the number of reversed spins is approximately
given by the equation
\begin{equation}
0.15 \ln(R)+0.25 R^{0.85}=\ln(8.1/\sqrt{B})\;.
\end{equation}
To take some typical values, for $B=2\;T$ and $7\;T$ the skyrmion is 
found to have $R\approx 7.8$  and 4.5 reversed spins, 
respectively, in good agreement with 
previous theoretical \cite {Fertig} and experimental \cite 
{Barrett,Schmeller,Goldberg} results.
Note also that spin-reversal produces a substantial correction to the 
$R=0$ QH energy; already for $R=8$, the energy is quite close to
$E_1(R=\infty)$.

\subsection{Skyrmions near $\nu=1/3$ and $\nu=1/5$}

Now we come to the skyrmion excitations near the $\nu=1/3$ state.
We first set the Zeeman splitting to zero.
At precisely $\nu=1/3$, the system is equivalent to $\nu^*=1$ of
composite fermions, and is thus expected to be fully polarized, as
found in numerical studies. The QH-skyrmion 
$N_{\phi}=3(N-1)+1$ maps into $N$  fermions at $N_{\phi}^*=N$,
and the QP-skyrmion $N_{\phi}=3(N-1)-1$
maps into $N$ fermions at $N_{\phi}^*=N-2$.

We have numerically studied a system of six electrons at 
$N_{\phi}=14$ and 16. The size of the Hilbert space can be reduced 
drastically by
confining to the sector with $L_z=S_z=0$, but is still too large for
exact diagonalization. (For $N_{\phi}=16$, the total
number of states in this sector is 16,004.) We have obtained several
low-energy states by Lanczos technique.
The CF analogy predicts that the low-energy states of the QP-skyrmion 
have quantum numbers $(L,S)=(0,0),\;(1,1),$ and (2,2), and those 
of the QH-skyrmion occur at   
$(L,S)=(0,0),\;(1,1),$ (2,2), and (3,3). Our results confirm this:
the lowest energy states with spins $S=0$, 1, 2 (and also 3 for 
$N_{\phi}=16$) indeed satisfy the property $L=S$, with energy 
increasing with $S$. 

To further ensure the validity of the CF mapping, we construct 
CF wave functions for the skyrmion at each $(L,S)$.
It was shown in Ref. \cite{WJ} that at small Zeeman energies, the
composite fermions (i.e., fermions at $N_{\phi}^*$) cannot be taken as
non-interacting. We consider two types of interactions, Coulomb and
hard-core, between them. The `hard-core' [`Coulomb'] trial wave function 
for composite fermions is given by Jastrow factor ($\Phi_1^2$ for the
spherical geometry) times the eigenstate of the 
hard-core [Coulomb] interaction for the $\nu=1$ skyrmion.
(Multiplication by $\Phi_1^2$ does not alter the $L$ and $S$ quantum
numbers.) 
As shown in Tables I and II, the Coulomb energies of these 
states  deviate from the exact Coulomb energy by $\sim$ 0.1\% or less, 
establishing that the $\nu=1/3$ skyrmion of electrons is indeed the 
$\nu=1$ skyrmion of composite fermions.

We note that there is no exact symmetry 
relating the QP- and
QH-skyrmions of the $\nu=1/3$ state.  However, they are 
related by particle-hole symmetry when viewed in terms of composite
fermions, clarifying the physics underlying the similarity between their
low energy spectra.

For large $N$, we use  $\Phi_1^2\Psi_{SK}(R)$ for the 
hard-core wave function for the 1/3-QH-skyrmion.
Its energy is shown in
Fig.~2 for several values of $N$.  For $N=50$:
\begin{equation}
E_{1/3}^{(0)}(R)/(e^2/\epsilon l)=0.069+0.024\exp(-0.38 R^{0.72})\;.
\label{nu=1/3}
\end{equation}
As before, adding the Zeeman term, expressing the energy in Kelvins
and $B$ in Tesla, we get:
\begin{equation}
E_{1/3}(R)=[3.45+1.20\exp(-0.38 R^{0.72})]\sqrt{B}+0.30 BR\;.
\end{equation}
The magnetic field must be less than $\approx 1.6\;T$ to produce any
spin reversal. The number of reversed spins can be determined 
straightforwardly by minimizing $E_{1/3}(R)$.

The energy of the skyrmion near $\nu=1/5$ is estimated for GaAs to
be (in K) 
\begin{equation}
E_{1/5}(R)=[1.7 + 0.37  \exp(-0.46 R^{0.75})]\sqrt{B} + 0.30 B R\;,
\end{equation}
from the wave function $\Phi_1^4\Psi_{SK}(R)$ for $N=40$ particles.
For $B$ greater than $\approx 0.21\;T$, the usual
fully polarized quasihole has the lowest energy, and the skyrmion
physics is not relevant.

These results show that the Zeeman energy is much more 
efficient in reducing the skyrmion size near $\nu=1/3$ and 1/5. 
This is not surprising, since the inter-CF interaction at $\nu^*=1$
is much weaker than inter-electron interaction at $\nu=1$.

The energy of the {\em QP}-skyrmion is harder to compute for
large $N$. We only make two observations here.
(i) The energies \cite {Girvin} of the QH and
QP-skyrmions are equal at $\nu=1$. This may suggest that
the energies of the two skyrmions should be equal also at $\nu=1/3$.
This is not the case. (If it were, the gap to creating a QP-QH
skyrmion pair at $\nu=1/3$, with zero Zeeman energy, would be $\sim
0.14 e^2/\epsilon l$, inconsistent with other estimates  of $ \sim 0.024
e^2/\epsilon l$ \cite 
{Sondhi,Moon}. For six electrons, the gap is $0.010 e^2/\epsilon l$.) 
The asymmetry can be 
understood as resulting from the fact that the vortices are attached 
only to electrons for both the QH- and QP-skyrmions.
(ii) The range over which the QP-skyrmion energy extends,
$E(0)-E(\infty)$, is estimated to be smaller than that for the
QH-skyrmion \cite {Moon}. (For $N=6$, it is
0.011 $e^2/\epsilon l$ for the QP-skyrmion as opposed to
0.017 $e^2/\epsilon l$ for the QH-skyrmion.) Therefore, any spin
reversal on the QP side is expected to require even lower
magnetic fields.

\subsection{Conclusion}

{\em A priori}, one might expect skyrmions near all odd integer  and
the corresponding fractional filling factors.
Experimental results of Schmeller {\em et al.} \cite {Schmeller} show 
that the low-energy spectrum at $\nu=3$,
5, ... does not contain any skyrmion-like states, in accordance with
the theoretical expectation \cite {JW}, also ruling out
any skyrmion-like structure near 
3/5, 3/7, 5/7, {\em etc}. This leaves 
the possibility of skyrmions only
near $\nu=1/(2m+1)$. While their observation is 
in principle possible at $\nu=1/3$ and 1/5, it would require
extremely low density GaAs samples.

Before closing, we note that we have not included above the 
effects of finite thickness and LL mixing, which are known to alter 
the QP and QH energies by as much as 50\%. We expect that these 
should lead to a more
or less $R$ independent shift in the finite-size skyrmion energy,
and therefore not change the above estimates significantly.
This work was supported in part by the National Science Foundation under 
Grant No. DMR93-18739.

\begin{table}
\begin{tabular}{|c|c|c|c|} \hline \hline
$(L,S)$ & exact Coulomb eigenstate &
{`hard-core' trial state} &
{`Coulomb' trial state}     \\ \hline \hline
(0,0)&-0.4620 & -0.4617 &  -0.4618  \\ \hline
(1,1)&-0.4616 & -0.4613 &  -0.4613  \\ \hline
(2,2)&-0.4604 & -0.4600 &  -0.4600  \\ \hline
\hline
\end{tabular}
\vspace{5mm}
\caption{Energies of the exact Coulomb eigenstates and of the `hard-core'
and `Coulomb' trial wave functions, explained in the text
for the $\nu=1/3$-QP-skyrmion with $N=6$ electrons at $N_{\phi}=14$.
The energies are given in units of $e^2/\epsilon l$,where $\epsilon$
is the background dielectric constant and $l=\protect\sqrt{\hbar c/eB}$
is the magnetic length, and include the interaction with the
positively charged neutralizing background.}
\end{table}

\begin{table}
\begin{tabular}{|c|c|c|c|} \hline \hline
$(L,S)$ & exact Coulomb eigenstate &
{`hard-core' trial state} &
{`Coulomb' trial state}     \\ \hline \hline
(0,0)&-0.4367 & -0.4362 &  -0.4362  \\ \hline
(1,1)&-0.4364 & -0.4359 &  -0.4360  \\ \hline
(2,2)&-0.4357 & -0.4354 &  -0.4353  \\ \hline
(3,3)&-0.4337 & -0.4333 &  -0.4333  \\ \hline
\hline
\end{tabular}
\vspace{5mm}
\caption{Same as in Table I for the QH-skyrmion with $N=6$ electrons
at $N_{\phi}=16$.}
\end{table}

\vspace{2cm}

\begin{figure}
\centerline{\psfig{file=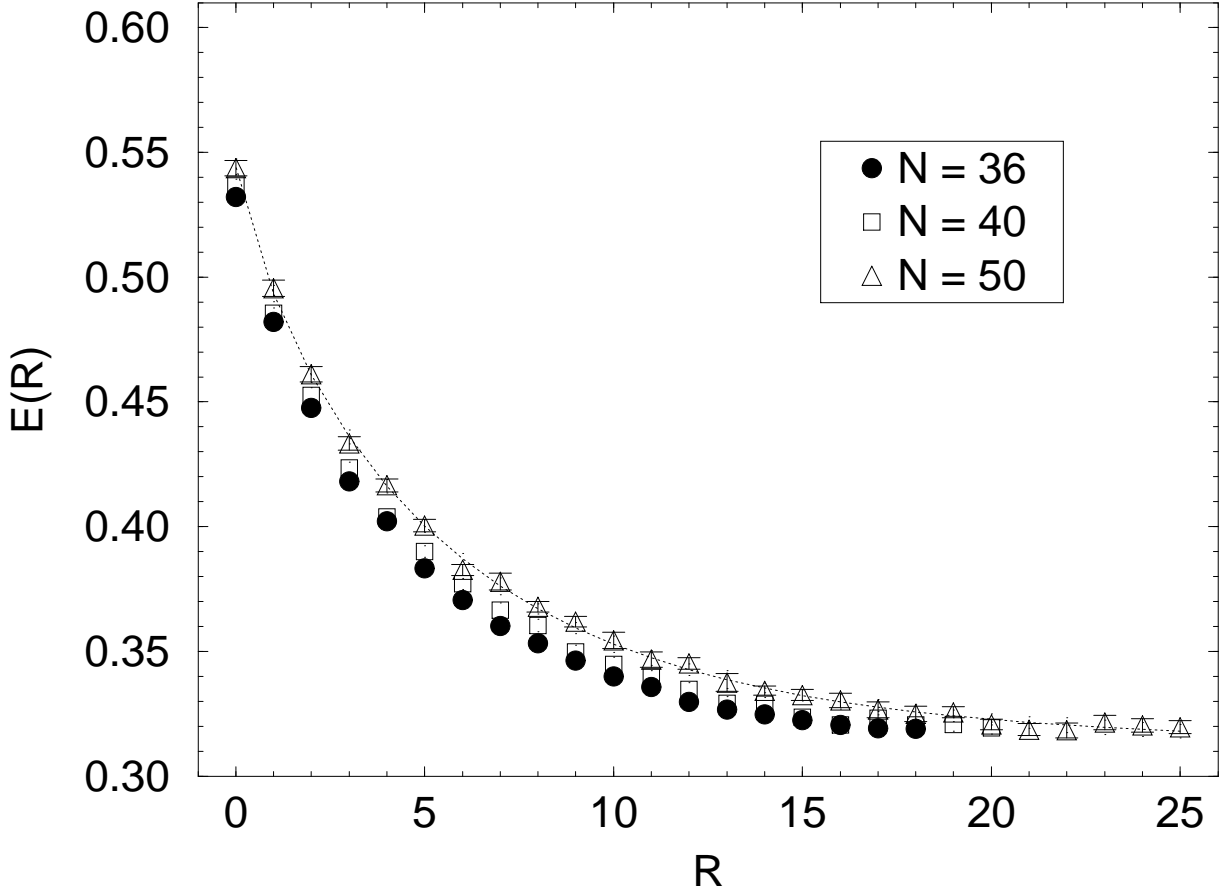,width=8.0in,angle=-90}}
\caption{Energy of the $\nu=1$ skyrmion in units of $e^2/\epsilon l$ for
several systems. The energy is measured relative to the fully
polarized $\nu=1$ state. The error bars are shown only for the $N=50$
electron system, and the dashed line plots Eq.~(\protect\ref{nu=1}).}
\label{fig:imp1}
\end{figure}

\begin{figure}
\centerline{\psfig{file=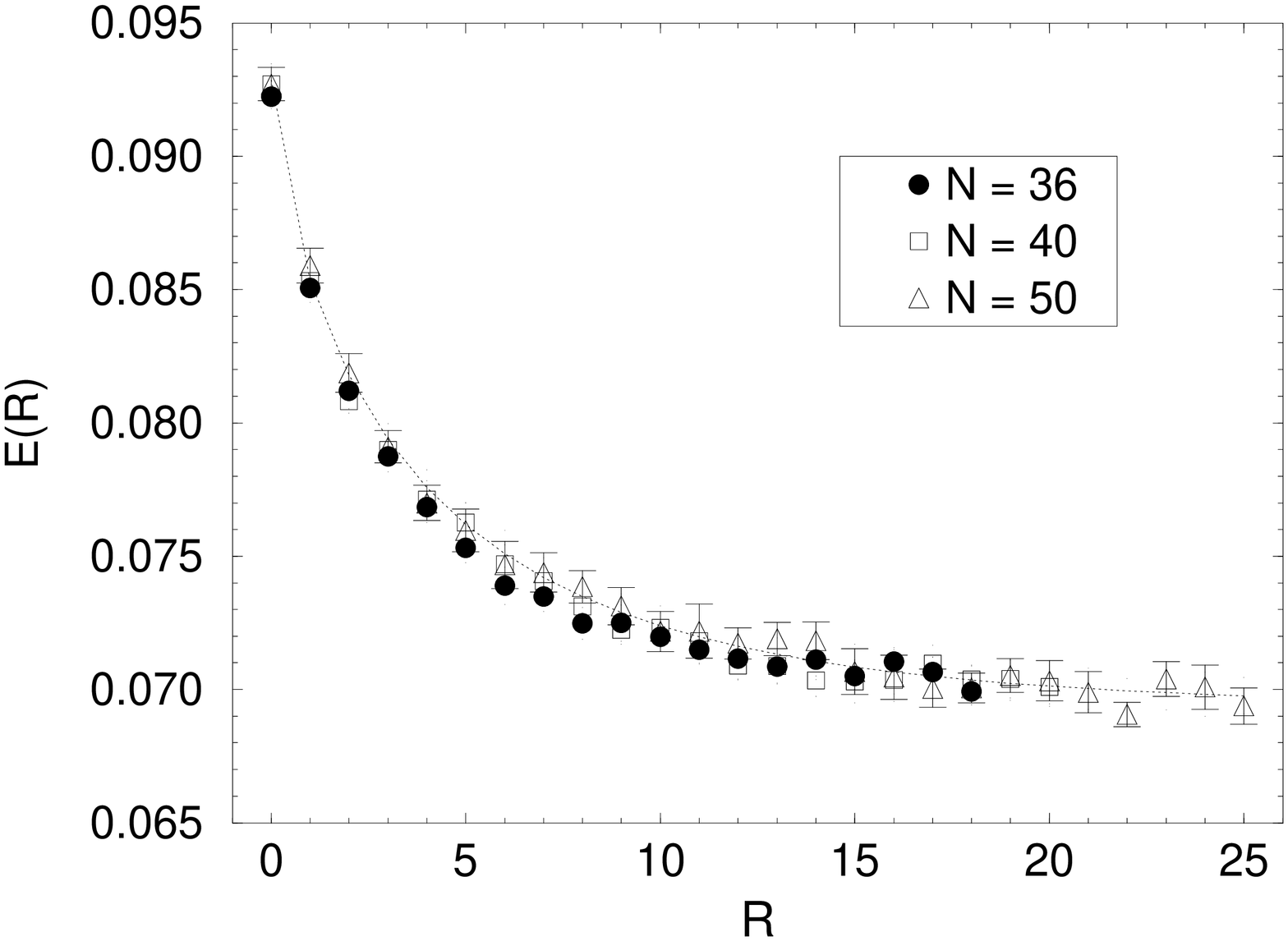,width=8.0in,angle=-90}}
\caption{Energy of the $\nu=1/3$ skyrmion measured relative to the fully
polarized 1/3 state. The error bars are shown only for the $N=50$
electron system, and the dashed line is a plot of Eq.~(\protect\ref{nu=1/3}).}
\label{fig:imp2}
\end{figure}


\begin{thebibliography}{99}


\bibitem{Tsui} D.C. Tsui, H.L. Stormer, and A.C. Gossard, Phys. Rev.
Lett. {\bf 48}, 1559 (1982).

\bibitem{Jain} J.K. Jain, Phys. Rev. Lett. {\bf 63}, 199 (1989);
Phys. Rev. B {\bf 41}, 7653 (1990); Science {\bf 266}, 1199 (1994).

\bibitem{Dev} G. Dev and J.K. Jain, Phys. Rev. Lett. {\bf 69}, 2843 (1992);
R.K. Kamilla, X.G. Wu, and J.K. Jain, Phys. Rev. Lett. {\bf 76}, 1332
(1996); X.G. Wu and J.K. Jain, Phys. Rev. B {\bf 51}, 
1752 (1995); N. Bonesteel,  Phys. Rev. B {\bf 51}, 9917 (1995).

\bibitem{HLR} B.I. Halperin, P.A. Lee, and N. Read, Phys. Rev. B {\bf
47} 7312 (1993); R.R. Du {\em et al.}, 
Phys. Rev. Lett. {\bf 70}, 2944 (1993); W. Kang {\em et al.}, 
{\em ibid.} {\bf 71}, 3850 (1993); V.J. Goldman {\em et al.}, 
{\em ibid.} {\bf 72}, 2065 (1994); R.L. Willett {\em et al.}, {\em ibid.}
{\bf 71}, 3846 (1993).


\bibitem{Du} R.R. Du, A.S. Yeh, H.L. Stormer, D.C. Tsui, L.N.
Pfeiffer, and K.W. West, Phys. Rev. Lett. {\bf 75}, 3926 (1995).

\bibitem{Wu} X.G. Wu, G. Dev, and J.K. Jain, Phys. Rev. Lett. {\bf
71}, 153 (1993).

\bibitem{Eisenstein} 
J.P. Eisenstein {\em et al.}, Phys. Rev. Lett. {\bf 61}, 997 (1987);
{\em ibid.} {\bf 62}, 1540 (1989); Phys. Rev. B {\bf 41}, 7910 (1990);
R.G. Clark {\em et al.}, Phys. Rev. Lett. {\bf 62}, 1536 (1989);
L. W. Engel {\em et al.},  Phys. Rev. B {\bf 45}, 3418 (1992).

\bibitem{Laughlin} R.B. Laughlin, Phys. Rev. Lett. {\bf 50}, 1395
(1983).

\bibitem{Nakajima} T. Nakajima and H. Aoki, Phys. Rev. Lett. {\bf 73},
3568 (1994).

\bibitem{Haldane} F.D.M. Haldane, Phys. Rev. Lett. {\bf 51}, 605 (1983);
T.T. Wu and C.N. Yang, Nucl. Phys. B {\bf 107}, 365 (1976).

\bibitem{Rezayi} E.H. Rezayi, Phys. Rev. B {\bf 43}, 5944 (1991); {\em
ibid.}, {\bf 36}, 5454 (1987).

\bibitem{Kane} D.H. Lee and C.L. Kane, Phys. Rev. Lett. {\bf 64}, 1313
(1990).

\bibitem{Sondhi} S.L. Sondhi, A. Karlhede, S.A. Kivelson, and E.H. Rezayi,
Phys Rev B {\bf 47}, 16419 (1993).


\bibitem{JW} J.K. Jain and X.G. Wu, Phys. Rev. B {\bf 49}, 5085
(1994).

\bibitem{He} S. He and X.C. Xie, Phys. Rev. B {\bf 53}, 1046 (1996).

\bibitem{MacDonald} A.H. MacDonald, H.A. Fertig, and L. Brey,
SISSA preprint cond-mat/9510080.

\bibitem{Girvin} The energies quoted here are for the 
`neutral' quasiparticle and
quasihole, in the convention of A.H. MacDonald and S.M. Girvin, Phys.
Rev. B {\bf 34}, 5639 (1986).

\bibitem{Fertig} H.A. Fertig, L. Brey, R. Cote, and A.H. MacDonald,
Phys. Rev. B {\bf 50}, 11018 (1994).

\bibitem{Barrett} S.E. Barrett, R. Tycko, L.N. Pfeiffer, and K.W.
West, Phys. Rev. Lett. {\bf 74}, 5112 (1995).

\bibitem{Schmeller} A. Schmeller, J.P. Eisenstein, L.N. Pfeiffer, and
K.W. West, Phys. Rev. Lett. {\bf 75}, 4290 (1995).

\bibitem{Goldberg} E.H. Aifer, B.B. Goldberg, and D.A. Broido, Phys. Rev.
Lett. {\bf 76}, 680 (1996). 

\bibitem{comm2} The $B$ here is the total magnetic field, while
the one that determines the magnetic length is the
normal component. We will assume here that the two are the same. The
analysis can be trivially generalized for tilted fields.

\bibitem{WJ} X.G. Wu and J.K. Jain, Phys. Rev. B {\bf 49}, 7515 (1994).

\bibitem{Moon} K. Moon {\em et al.} Phys. Rev. B {\bf 51}, 5138
(1995).

\end{thebibliography}
\end{document}